\documentstyle[aps,preprint,12pt]{revtex}

\begin{document}
\begin{sl}
\title{ON THE GENERAL COVARIANCE IN THE BOHMIAN QUANTUM GRAVITY}
\author{FATIMAH SHOJAI\footnote{Email: FATIMAH@NETWARE2.IPM.AC.IR}$^{,1,2}$
\& MEHDI GOLSHANI\footnote{Fax: 98-21-8036317}$^{,1,2}$}
\address{$^1$Department of Physics, Sharif University of Technology\\P.O.Box 11365-9161 Tehran, IRAN.}
\address{$^2$Institute for Studies in Theoretical Physics and Mathematics,\\P.O.Box 19395-5531, Tehran, IRAN.}
\date{\today}
\maketitle
\begin{abstract}
{\it It is shown explicitly that in the framework of Bohmian quantum gravity, the
equations of motion of the space-time metric are Einstein's equations plus some 
quantum corrections. It is observed that these corrections are not covariant. So
that in the framework of Bohmian quantum gravity the general covariance principle
breaks down at the individual level. This principle is restored at the statistical level.}
\end{abstract}
\section{INTRODUCTION}
\par
In the de-Broglie--Bohm interpretation of quantum mechanics\cite{Bohm}, the quantum effects
are described by the quantum potential. It has at least two peculiar properties,
it is non-local, and it is able 
to break down the classical symmetries for an individual process while they remain valid
statistically. For example, the above-mentioned problem appears for the Lorentz
symmetry. So we have velocities greater than light's velocity for an individual
phenomenon.\cite{Bohm,Slm}
\par
General relativity is invariant under the general coordinate transformations.
It is not clear whether this symmetry preserved after quantization or not. In the Copenhagen
quantum gravity, the constraints related to this symmetry (one on Hamiltonian and
three on momenta) appear weakly (in the terminology of Dirac's 
canonical quantization procedure). For the disscusion about the 
governing algebra, it is necessary to represent the operators as well-defined 
ones and then one obtains the commutation relations between them.
\par
The viewpoint of the Copenhagen approach is, as usuall, statistical. For an 
individual description we must use Bohm's theory. In this theory, the
quantum corrections to the constraints are represented by terms involving
the quantum potential. But, the statistical results of this interpretation
is the same as those of the Copenhagen quantum mechanics.
\par
Essentially, in the Bohmian quantum gravity, Einstein's equations (which
are the equations of motion of the space-time metric) would be modified by some 
expressions containing the quantum potential. In this paper we shall derive these 
modified Einstein's equations. As a result, the general covariance principle 
would be broken down at the individual level, but it remains valid statistically.
\section{BOHMIAN QUANTUM GEOMETRODYNAMICS}
\par
In this section we discuss the quantum geometrodynamics via the de-Broglie--Bohm 
interpretation. The Wheeler-De Witt (WDW) approach\cite{WDW} for quantization of gravity is based on the 3+1
decomposition of space-time (ADM decomposition). This splitting is necessary 
because we use the canonical quantization of the Hamiltonian, and the Hamiltonian 
formulation is not covariant. In the ADM decomposition the space-time is 
decomposed into spatial slices which are labled by a time variable $t$ ($\Sigma_t$).
In this manner, the metric is specified by lapse ($N$) and shift ($N_i$) functions
and the induced metric ($h_{ij}$). The line-element is:
\begin{equation}
ds^2=(Ndt)^2-h_{ij}(N^idt+dx^i)(N^jdt+dx^j)
\end{equation}
The extrinsic curvature of the spatial surfaces $\Sigma_t$'s are given by:
\begin{equation}
K_{ij}=\frac{1}{2N}\left [ {\cal D}_jN_i+{\cal D}_iN_j-\dot{h}_{ij}\right ]
\end{equation}
where the covariant derivative ${\cal D}_i$ is defined with respect to 
$h_{ij}$ metric and the dot above letters represents time derivative.
The Einstein-Hilbert action can be written as:
\[ {\cal A}=\frac{1}{16\pi G}\int d^4x \sqrt{-g}{\cal R}\]
\begin{equation}
=-\frac{1}{16 \pi G}\int d^3x \sqrt{h}N\left [ K^2-K_{ij}K^{ij}
+\ ^{(3)}{\cal R}\right ] + {\em surface\ terms}.
\end{equation}
where $K=trace(K_{ij})$, $h=det(h_{ij})$ and $^{(3)}{\cal R}$ is the 
three-dimmensional
curvature. If one defines the conjugate momenta as usuall, we have:
\begin{equation}
\pi=\frac{\delta {\cal L}}{\delta \dot{N}}=0
\end{equation}
\begin{equation}
\pi^i=\frac{\delta {\cal L}}{\delta \dot{N_i}}=0
\end{equation}
\begin{equation}
\pi^{ij}=\frac{\delta {\cal L}}{\delta \dot{h}_{ij}}=\frac{\sqrt{h}}{16\pi G}
\left ( K^{ij}-Kh^{ij}\right )
\end{equation}
It is concluded that the $N$ and $N_i$ functions aren't dynamical and $h_{ij}$
is the only dynamical degree of freedom. The equations (4) and (5) are 
primary constraints. The Hamiltonian is:
\begin{equation}
H=\int d^3x (N{\cal H}_G+N^i{\cal H}_i)
\end{equation}
where
\begin{eqnarray}
&{\cal H}&_G=16 \pi G G_{ijkl}\pi^{ij}\pi^{kl}-\frac{\sqrt{h}\ ^{(3)}{\cal R}}{16\pi G}\\
&{\cal H}&^i=-\frac{1}{8\pi G}{\cal D}_j\pi^{ij}\\
&G&_{ijkl}=\frac{1}{2\sqrt{h}}\left ( h_{ik}h_{jl}+h_{il}h_{jk}-h_{ij}h_{kl}\right )
\end{eqnarray}
Since the first constraints must be satisfied at all times, we must have:
\begin{equation}
\dot{\pi}=-\{H,\pi\}=\frac{\delta H}{\delta N}=0\Longrightarrow {\cal H}_G=0
\end{equation}
\begin{equation}
\dot{\pi_i}=-\{H,\pi_i\}=\frac{\delta H}{\delta N^i}=0\Longrightarrow {\cal H}_i=0
\end{equation}
These equalities are the secondary constraints, which are satisfied at all times
because of the specific form of the Hamiltonian. Thus we have no new constraints 
anymore.
In the canonical quantization, the relations (11) and (12) limit the physical state domain.
In other words, among all the states in the Hilbert space, some special ones are physical.
We distinguish these states by applying the constraints (11) and (12) weakly, i.e.:
\begin{eqnarray}
&\hat{{\cal H}}&_G\Psi[h_{ij}]=0\\
&\hat{{\cal H}}&_i\Psi[h_{ij}]=0
\end{eqnarray}
where $\Psi$ is the physical wavefunction of the universe.
\par
According to the canonical quantization process, the $\hat{{\cal H}}_i$
and $\hat{{\cal H}}_G$ operators can be obtained by substituding the
canonical momenta $\pi^{ij}$ with
$-i\frac{\delta}{\delta h_{ij}}$ in ${\cal H}^i$  and ${\cal H}_G$ which
are the three and one-dimmensional diffeomorphism
generators on the space-like surfaces and time-like direction, respectively.
Then the wave-function of the universe must be annihilated by these generators.
The constraint $\hat{{\cal H}}_G\Psi=0$ (the Hamiltonian constraint) is the 
WDW equation. Its extension to the case in which the matter field $\phi$ exist, is:
\begin{equation}
\left [ 16\pi G h^{-q}\frac{\delta}{\delta h_{ij}}h^qG_{ijkl}
\frac{\delta}{\delta h_{kl}}+\frac{\sqrt{h}}{16\pi G}\ ^{(3)}{\cal R}
-{\cal T}_0^0(\phi, -i\partial/\partial \phi)\right ] \Psi[h_{ij},\phi]=0
\end{equation}
where ${\cal T}_{\mu \nu}$ is the matter field energy-momentum tensor, and $q$ 
is the ordering parameter. This equation and:
\begin{equation}
\hat{{\cal H}}_i\Psi=0 \Longrightarrow -\frac{1}{8\pi G}{\cal D}_j \frac{\delta \Psi}{\delta h_{ij}}
+\partial ^i \phi \frac{\delta \Psi}{\delta \phi}=0
\end{equation}
specify the wavefunction of the universe.
\par
The WDW equation has the following points:
\begin{itemize}
\item The time parameter which defines the foilation of the space-time, doesn't
appear in it. (the so-called time-problem in quantum gravity)
\item A different ordering of factors leads to a different result.
\item In practice, for solving the WDW equation, instead of using an infinite-dimmensional
superspace, we must limit ourselves to a mini-superspace in which some of the degrees
of freedom are non-frozen.
\item It is necessary for the wave-function to be square-integrable, in order
to have a probabilistic interpretation for it. But this is not possible
for all cases, because a precise definition of the inner product is not 
known in quantum gravity.
\item The WDW equation contains a multiplication of 
two functional derivatives which are calculated at the same point. Then applying 
the WDW operator on $\Psi$, we have a multiplication of two delta functions at one point. Therefore
we must regularize the kinetic term of the WDW equation as:
\begin{equation}
h^{-q}\frac{\delta}{\delta h_{ij}(\vec{x})}h^q G_{ijkl}\frac{\delta \Psi}{\delta h_{kl}(\vec{x})}
\longrightarrow \underbrace{h^{-q}\frac{\delta}{\delta h_{ij}(\vec{x})}
h^q \tilde{G}_{ijkl}(\vec{x},\vec{x}';t)\frac{\delta \Psi}{\delta h_{kl}(\vec{x}')}}_{\Delta_{reg.}}
\end{equation}
where $\lim_{t\rightarrow 0}\tilde{G}_{ijkl}(\vec{x},\vec{x}',t)=G_{ijkl}\delta (\vec{x}-\vec{x}')$
and $\tilde{G}_{ijkl}$ satisfies the heat equation (heat kernel).
The physical wave-functions must be annihilated by constraints. Thus:
\begin{equation}
\Psi_{physical}=\delta({\cal H}_G^{reg.})\Psi [h_{ij}]=
\int DM(x) e^{i\int d^3x M(x){\cal H}_G^{reg.}}\Psi [h_{ij}]
\end{equation}
\item In the classical limit, we have:
\begin{equation}
\{{\cal H}_i,H\}= \{{\cal H}_G,H\}=0 
\end{equation}
where $\{,\}$ represents the poisson bracket. This means that ${\cal H}_i$ and 
${\cal H}_G$ form  a closed algebra, and no new constriant appears. But this 
fact is problematic at the quantum level.
\end{itemize}
\par
Now, we use the canonical transformation $\Psi(h_{ij})=\Gamma(h_{ij})e^{iS(h_{ij})}$
in the equations (15) and (16), ignoring the matter fields for simplicity. Equating
the real and imaginary parts of the equation (15), one gets:
\begin{eqnarray}
&&16\pi G \tilde{G}_{ijkl}\frac{\delta S}{\delta h_{ij}}\frac{\delta S}{\delta h_{kl}}
-\frac{\sqrt{h}}{16\pi G}(^{(3)}{\cal R}-Q_G)=0\\
&&\frac{\delta }{\delta h_{ij}}\left [  h^q \tilde{G}_{ijkl}\frac{\delta S}{\delta h_{kl}}\Gamma^2\right ]=0
\end{eqnarray}
where $Q_G$ is the quantum potential of the gravitational field:
\begin{equation}
Q_G(h_{ij})=-\frac{1}{\sqrt{h}\Gamma}\left ( \tilde{G}_{ijkl}\frac{\delta^2\Gamma}{\delta h_{ij}
\delta h_{kl}}+h^{-q}\frac{\delta h^q\tilde{G}_{ijkl}}{\delta h_{ij}}\frac{\delta \Gamma}{\delta h_{kl}}\right )
\end{equation}
Also equation (16) reads as:
\begin{eqnarray}
{\cal D}_j\frac{\delta \Gamma}{\delta h_{ij}}=0\\
{\cal D}_j\frac{\delta S}{\delta h_{ij}}=0
\end{eqnarray}
Equation (20) is a modified Hamilton-Jacobi equation. It indicates that the only difference between
classical and quantum universes is the existance of the quantum potential in the latter.
Equation (21) shows the conservation of the probability in the superspace. In addition, in the 
de-Broglie--Bohm theory, the guiding formula defines the momenta corresponding to 
the coordinates. Then, for the dynamical coordinates $h_{ij}$ we have:
\begin{equation}
\pi^{kl}=\frac{\delta S}{\delta h_{kl}}=\frac{\sqrt{h}}{16\pi G}(K^{kl}-h^{kl}K)
\end{equation}
Here it must be noted that the Hamilton-Jacobi equation indicates the dynamical properties. So that 
one can deal with the Hamilton equations of motion instead of equations (20)-(25).
So we have, equivalently, the following relations:
\begin{equation}
\dot{\pi}_{ij}=\{\pi_{ij},\tilde{H}\}
\end{equation}
\begin{equation}
\dot{h}_{ij}=\{h_{ij},\tilde{H}\}
\end{equation}
where
\begin{eqnarray}
&\tilde{H}&=\int d^3x (N\tilde{{\cal H}}_G+N^i\tilde{{\cal H}}_i)\\
&\tilde{{\cal H}}&^i=-\frac{1}{8\pi G}{\cal D}_j\pi^{ij}\\
&\tilde{{\cal H}}&_G=16\pi G \tilde{G}_{ijkl}\pi^{ij}\pi^{kl}-\frac{\sqrt{h}}{16\pi G}(^{(3)}{\cal R}
-Q_G)
\end{eqnarray}
The de-Broglie--Bohm approach has the following advantages:
\begin{itemize}
\item Although the time parameter does not appear in the wave-function, it emerges from the guiding
formula, naturally.
\item In this theory the role of the wave-function is different from
the Copenhagen quantum mechanics. Its phase indicates the evolution of the dynamical variables
according to the guiding formula. Its amplitude charactrizes the quantum potential which
includes all the quantum effects. These two specifications of the wave-function
have appeared for the individual processes and the wave-function may be non-normalized.
The other aspect of the wave-function appears at the statistical level. The square of its amplitude
has the probability interpretation as in the Copenhagen quantum mechanics, and
therefore it is necessary to be normalized.
\item An important problem, with which we are concerned, is the role of the
${\cal H}_i$ and ${\cal H}_G$ constraints at the quantum level. We shall deal with
this point in the following section.
\end{itemize}
\section{QUANTUM POTENTIAL AND THE GENERAL COVARIANCE}
\par
In this section, we first review some viewpoints about the role of the constraints
in quantum gravity:
\begin{itemize}
\item Gilkman \cite{Gilkman} has used the de-Broglie--Bohm approach. Because of
existance of the quantum potential term in ${\cal H}_G$, he has shown that the
constraints' algebra is not closed. Therefore in order for the constraints to remain
valid at all times, one obtains some new constraints, etc. Consequently,
he believes that the symmetry given by the four dimmensional diffeomorfism
dosen't exist for individual processes, after quantization.
In his view, this point is related either to the existance of a
minimal length in quantum gravity, or probably to the use of the ADM
decomposition for quantization.
\item Shtanov \cite{Shtanov} has pointed out the problem of constraints. He believes 
that in the classical mechanics the choice of a Lagrange multiplier does not have any effect
on the physical solution. But in quantum gravity, using the de-Broglie--Bohm
approach, the situation is different.
From the guiding formula for $g_{\mu \nu}$ one sees that the role of $N_i$ in
the quantum dynamics is the same as in classical dynamics. But the $N$ function
plays different roles in the classical and quntum domains. After quantization,
a non-local function (quantum potential) appears in $H_G$ and causes the
physical charactristics of $g_{\mu \nu}$ to depend on $N$, in the general case. 
In the classical limit, one ignores the quantum potetial in comparison with the 
classical potentials. In this limit, the dependence is removed.
\par
Shtanov concludes that the quantum dynamics of gravity breaks down the
foliation-invariance of the classical general relativity. This is a result of the
quantum non-locality. Of course, it must be noted that the
foliation-invariance breaking only exists for the individual processes, 
according to the de-Broglie--Bohm theory. Because in this theory, it is not
necessary for the dynamics of an individual system to follow all the
statistical invariances.
\item Horiguchi et. al.\cite{Horiguchi} have first regularized the WDW eqation
and then normalized it, by preserving the three-dimmensional general covariance.
Therefore, the momenta-constraints algebra doesn't leads to some new
constraints (anomali freedom of momenta constraints). Then, they have
considered the Hamiltonian constraint algebra and indicated that an anomalous
term may appear from the commutation between $\Delta _{reg.}$ and $\sqrt{h}\ ^{(3)}{\cal R}$.
If one sets this commutation relation equal to zero, a new constraint would 
result besides the WDW equation. 
\item Blaut et. al.\cite{Blaut} have obtained the regularized WDW equation with
an anomaly free condition for constraint algebra. They have observed that this condition 
is satisfied only for a specific subset of the wave-functions. This subset contains
the wave-functions that are functions of 3-scalar densities. Then, they have
used the quantum potential approach and shown that quantum gravity has less 
symmetry than classical gravity. The definition of $N$ in the Hamiltonian is not 
free and it is fixed by the Hamiltonian constraint. In other words, the general 
covariance breaks down because of the non-existance of time translational 
symmetry at the individual level. This fact must have some physical effects at 
the Planck scale. In the classical limit, where we ignore the quantum effects,
the time translational symmetry would be restored.
\end{itemize}
\par
From the foregoing discussion, we can deduce an important result:
{\it It is possible to find some wave-function for which the constraints' algebra
is satisfied, using the regularized WDW equation}. Thus, from the Bohmian point of 
view, the constraints can be met statistically, but not necessarily individually.
\par
In this paper, our aim is to discuss explicitly the break down of the constraints
in confirmation of the above result. It is a well-known fact that four equations of
the Einstein's equations are constraints on the extrinsic curvature ($K^{ij}$) and the
3-space metric, and the remining equations represent the time evolution of the 
3-space metric. This point results from the fact that some of the Reimann
tensor components depends only on the extrinsic curvature and the 
3-space  intrinsic curvature.
\par
From the Gauss-Codadzi equations \cite{GC} we have:
\begin{eqnarray}
&{\cal R}&^0_{ijk}={\cal D}_iK_{jk}-{\cal D}_jK_{ik}\\
&{\cal R}&^m_{ijk}=\ ^{(3)}{\cal R}^m_{ijk}+K_{jk}K^m_i-K_{ik}K^m_j
\end{eqnarray}
Using these, one obtains:
\begin{equation}
G^0_0=-\frac{1}{2}\left [ \ ^{(3)}{\cal R}+K^2-K_{ij}K^{ij}\right ]
\end{equation}
Thus, one of the Einstein's equations relates the extrinsic curvature of the space-like
slices and their intrinsic scalar curvature. Furthurmore, one other result from
the equations (31) and (32) is:
\begin{equation}
G^0_i={\cal D}_jK_i^j-{\cal D}_iK
\end{equation}
Therefore, these three Einstein's equations are constraints on the extinsic curvature
of the space-like surfaces.
\par
After quantization, the constraints (33) and (34) would appear in the form of 
equations (15)-(16) in the absence of the matter field or (20)-(24), from the Copenhagen or Bohmian points of view,
respectively. Starting from the Bohmian form, the relation (20) gives:
\begin{equation}
16\pi G \tilde{G}_{ijkl}\pi^{ij}\pi^{kl}-\frac{\sqrt{h}}{16\pi G}(\ ^{(3)}{\cal R}-Q_G)=0
\end{equation}
On using the equation (25) and the fact that $G_{ijkl}\pi^{kl}=K_{ij}/16\pi G$, we have:
\begin{equation}
(K^{ij}-Kh^{ij})K_{ij}-\ ^{(3)}{\cal R}+Q_G=0
\end{equation}
Thus:
\begin{equation}
G^0_0=-\frac{Q_G}{2}
\end{equation}
Therefore taking the quantum effects into consideration, the constraint $G^0_0=0$
would be corrected according to the relation (37). The other three constraints at
the quantum level can be obtained from equation (24):
\begin{equation}
{\cal D}_j\pi^{ij}=0
\end{equation}
Since ${\cal D}_jh^{ij}=0$, we have:
\begin{equation}
{\cal D}_jK^j_i-{\cal D}_iK=0
\end{equation}
Therefore:
\begin{equation}
G^0_i=0
\end{equation}
Thus the three-dimmensional diffeomorphism constraints are not changed when 
one takes the quantum effects into account. This fact can be also concluded
by observing the absence of the quantum potential in the relation (24).
\par
As we argued previously, the role of the momenta constraints for the
classical and quantum cases is the same. But the Hamiltonian constraint for
the quantum case is corrected by the quantum potential.
\par
Now, we must obtain the corrections of the dynamical Einstein's equations (i.e the
spatial components of $G_{\mu \nu}$). These are the governing equations on the metric
of the space-like surfaces $h_{ij}$ (i.e. the equations of motion of $h_{ij}$).
By noting to the relations (3) and (33), the Einstein-Hilbert action, in the ADM 
decomposition can be written as:
\begin{equation}
{\cal A}=\frac{1}{8\pi G}\int d^4x N\sqrt{h}G^0_0+{\em surface\ terms}
\end{equation}
By varying ${\cal A}$ with respect to $h_{mn}$, one obtains:
\begin{equation}
G^{mn}(x)=-2\int d^4x' N \frac{\delta}{\delta h_{mn}(x)}(\sqrt{h(x')}G_0^0(x'))
\end{equation}
So that if the Hamilton-Jacobi equation be multiplied by $N$ and varied with respect
to $h_{mn}$, the equation of motion of the 3-space metric would be obtained. Thus, 
substitution from (37) yields:
\begin{equation}
G^{mn}(x)=\int d^4x' N \frac{\delta}{\delta h_{mn}(x)}[\sqrt{h(x')}Q_G(x')]
\end{equation}
\par
This can also be deduced from the Hamilton-Jacobi equation. We know from the 
Hamilton-Jacobi formalism that by varying the Hamilton-Jacobi equation with
respect to any coordinate and using the guiding formula, the equation of motion 
of that coordinate would be obtained.
Therefore, varying the relation (35) and doing some algebra, leads
to the relation (43).
\par
Thus, the dynamical equations of the space-like surface metric are also corrected 
by the quantum potential.
\section{OBSERVATIONS}
\par
Now we are ready to discuss some important results:
\begin{itemize}
\item In the Bohmian quantum theory of gravity, the general covariance,
represented by $G^0_0=0$ and $G_i^0=0$ constraints, breaks down. This is because 
of the breaking of $G_0^0=0$ constraint for individual processes. The equations (37), (40)
and (42) are not covariant and involve the spatial and time-like components
differently.
The Break down of the general covariance principle is caused by the quantum potential
and shows that the equivalence principle is not valid for individual processes at 
quantum level, necessarily. Since in one sense, the equivalence principle
is in contradiction to Mach's principle (i.e. in a local inertial frame, the laws of 
motion are independent of the distant matter), the break down of the former may be a
step towards the latter. This point is discussed in the ref. \cite{SG} in detail.
\item According to the equations (37), (40) and (42), the modified Einstein's equations,
are functianals of $h_{ij}$ and the scalar $\Gamma$. This is suggesting that
it is probably necessary to use a scalar-tensor theory for the quantum
description of gravity. 
\item Although we started from a pure gravity field, in the general case with a matter
field, one can do in a similar way and obtain the modified Einstein's equations as:
\begin{eqnarray}
&G&_0^0=\kappa {\cal T}_0^0-\frac{Q_G+Q_M}{2}\\
&G&_i^0=\kappa {\cal T}_i^0\\
&G&^{mn}=\kappa {\cal T}^{mn}+\int d^4x' N \frac{\delta}{\delta h_{mn}}[\sqrt{h}(Q_G+Q_M)]
\end{eqnarray}
where $Q_M$ is the quantum potential of the matter resulted from the dependence of 
$\Gamma$ upon the matter field and is independent of the ordering parameter.
For example, for a scalar field, we have:
\begin{equation}
Q_M=-\frac{1}{h}\frac{1}{\Gamma}\frac{\delta^2\Gamma}{\delta \phi^2}
\end{equation}
\item In Bohm's theory, the Hamilton-Jacobi equation and the Newton's equation of
motion are respectively:
\begin{eqnarray}
&&\frac{\partial S}{\partial t}+\frac{|\vec{\nabla}S|^2}{2m}+V+Q=0\\
&&\frac{d\vec{p}}{dt}=-\vec{\nabla}(V+Q)
\end{eqnarray}
where $Q$ is the quantum potential. Therefore in the classical limit, it is necessary
that $Q$ be numerically negligible ($Q<<$ other energies in Hamilton-Jacobi equation)
and slowly varying (at least its first derivatives should be negligible: $\vec{\nabla}Q<<\vec{\nabla}V$).
Now, in the modified Einstein's equations, relations (44) and (46) are the Hamilton-Jacobi
equation and the modified equation of motion. Thus in the classical limit, we
ignore the second terms in the right hand side of these equations in comparison to 
the first term, and then the conventional Einstein's equations are obtained 
explicitly.
\item One way to preserve the general covariance principle for the individual 
processes is to see the quantum potential as an agent for the quantum force.
Just as in the nonrelativistic Bohm's theory we correct the Newton's second
law as the relation (49), one can correct the Einstein's equations similarly.
In a covariant way, one may write:
\begin{equation}
G^{\mu \nu}=\kappa {\cal T}^{\mu \nu }+F_Q^{\mu \nu}
\end{equation}
where $F_Q^{\mu \nu}$ is a second rank (quantum force) tensor, appeared at the
right hand side as a quantum source for gravity. 
\par
It must be pointed out that this equation is a new extension of the de-Broglie--Bohm
theory for gravity. Naturally, its results are not necessarily the same as the 
Copenhagen quantum gravity or even the conventional de-broglie--Bohm theory 
of gravity. But this extension must be such that in some special limit, the 
usuall quantum gravity would result.
\item From every aspect, the quantum potential is very important in quantum
gravity. In reference \cite{S}, it is shown that the existance of the matter 
quantum potential is equivalent to introduction of a conformal factor in the space-time
metric. This means that the quantal effects of the matter may be thought as 
some geometrical effects. This viewpoint about the quantum potential has many 
advantages. For example it is able to remove the cosmological singularities
in the early universe. \cite{S}
\end{itemize}
\end{sl}

\end{document}